\documentclass[12pt,preprint]{aastex}
\usepackage{natbib}
\usepackage{epsfig,graphicx}

\shorttitle{Velocity Dispersion of Excited H$_{2}$}
\shortauthors{S. Lacour et al.}

\received{2004 June 12}
\begin{document}


\title{Velocity Dispersion of the High Rotational Levels of H$_{2}$}


\author{S. Lacour\altaffilmark{1,2,3}, V. Ziskin\altaffilmark{2,1}, 
G. H{\' e}brard\altaffilmark{1,2}, C. Oliveira\altaffilmark{1}, \\
M. K. Andr{\' e}\altaffilmark{2,4}, 
R. Ferlet\altaffilmark{2},
and A. Vidal-Madjar\altaffilmark{2}}


\altaffiltext{1}{The Johns Hopkins University, Department of Physics
and Astronomy, 3400 North Charles Street, Baltimore, MD 21218, USA}
\altaffiltext{2}{Institut d'Astrophysique de Paris, CNRS, 98 bis
Boulevard Arago, F-75014 Paris, France} 
\altaffiltext{3}{Observatoire
de Paris/Meudon, LESIA, 5 Place Jules Janssen, F-92195 Meudon, France}
\altaffiltext{4}{AZimov association, 14 rue Roger Moutte, F-8270 St CYr, France}


\begin{abstract}   

We present a study of the high rotational bands ($J \geq 2$) of H$_2$
toward 4 early type galactic stars: HD\,73882, HD\,192639, HD\,206267,
and HD\,207538.  In each case, the velocity dispersion - characterized
by the spectrum fitting parameter $b$ - increases with the level of
excitation, a phenomenon that has previously been detected by the {\it
Copernicus} and {\it IMAPS} observatories. In particular, we show with
$4\sigma$ confidence that for HD\,192639 it is not possible to fit all
$J$ levels with a single $b$ value, and that higher $b$ values are
needed for the higher levels.  The amplitude of the line broadening,
which can be as high as 10 km\,s$^{-1}$, makes explanations such as
inhomogeneous spatial distribution unlikely.  We investigate a
mechanism in which the broadening is due to the molecules that are
rotationally excited through the excess energy acquired after their
formation on a grain (H$_2$-formation pumping). We show that different
dispersions would be a natural consequence of this mechanism. We note
however that such process would require a formation rate 10 times
higher then what was inferred from other observations.  In view of the
difficulty to account for the velocity dispersion as thermal
broadening ($T$ would be around 10\,000 K), we conclude then that we
are most certainly observing some highly turbulent warm layer
associated with the cold diffuse cloud. Embedded in a magnetic field,
it could be responsible for the high quantities of CH$^+$ measured in
the cold neutral medium.

\end{abstract}




\section{Introduction}

The population distribution of the different rotational levels ($J$)
of H$_2$ provides detailed information about diffuse and translucent
clouds. The kinetic temperature, derived from the column density
distribution of H$_2$ in the $J$ = 0, 1 and 2 levels, is usually about
80 K. On the other hand, the excitation temperature obtained from the
$J$ = 3 - 6 levels is typically of several hundred K.  The
measurement of these column densities is an important probe of the
physical conditions of the interstellar gas. Temperature, density, UV
radiation field and other parameters can be inferred from such
information, but a reliable model for the distribution of populations
is required.

There are three distinct mechanisms that likely determine the
population of the excited levels: ultraviolet pumping, H$_2$ formation
on grains, and high temperature collisional processes.  Of these, the
UV photoexcitation process \citep[as described
in][]{1976ApJ...203..132B} has been generally considered to be
dominant. In the case of $\zeta$ Oph \citep{1977ApJS...34..405B}, for
instance, good matches were obtained for both the H$_2$ population
distribution and the abundances of all known chemical species, except
CH$^+$.

Additional information, explored in this paper, can be obtained from
the presence of a measurable velocity dispersion which is an
increasing function of the rotational energy level ($J$). This effect
was first seen in several {\it Copernicus} observations, and was seen
in both the curve of growth $b$-value and in the line widths
\citep{1973ApJ...186L..23S,1974ApJS...28..373S}. More recently,
\citet{1997ApJ...477..265J} using {\it IMAPS} data (with a resolution
power of 120,000), showed, for the two main resolved components in the
line of sight towards $\zeta$ Ori A, a clear broadening of the H$_2$
lines increasing with rotational level.  Along that line of sight, at
least in the component showing the highest broadening, there is no
doubt that the effect is linked to the excitation process of the
H$_{2}$ molecule.

It is this interplay between rotational excitation and velocity
dispersion that we explore in this paper. Data from the {\it FUSE} FUV
satellite \citep{2000ApJ...538L...1M} offers new insights on this
issue.  The wavelength range covered by {\it FUSE} (905 to 1187 \AA)
contains more than twenty Lyman vibrational transitions, as well as
six Werner bands, providing a large number of transitions for each
rotational ground state and allowing measurements over a wide range of
oscillator strengths. In addition, the high sensitivity of the
instrument puts within observational reach some interesting lines of
sight with high extinction, such as HD\,73882, with $E(B-V)$ = 0.73
\citep{2000ApJ...538L..65S}. Although the spectral resolution does not
allow us to resolve the different absorption components, we will show
that saturation effects can also reveal the velocity dispersion of the
lines.

This paper is organized as follows. The observations and data
reduction are described in Section~\ref{sec:obs}.
Section~\ref{sec:targe} presents the H$_2$ analysis along the four
sightlines listed in Table~\ref{targets}. Calculations were done using
the curve of growth method (hereafter COG), and profile fitting
(hereafter PF). Both methods are explained and discussed.  In
Section~\ref{sec:conc}, we compare the observations with the
theoretical explanation of the excitation, and discuss the
consequences on the chemistry of the cloud. In the Appendix we
describe a model of H$_2$ formation and subsequent cooling, which
could explain the observed velocity dispersion.

\section{Observations \& Data Reduction}
\label{sec:obs}

The {\it FUSE} mission, its planning, and its in-orbit performance are
discussed by \citet{2000ApJ...538L...1M} and
\citet{2000ApJ...538L...7S}. Briefly, the {\it FUSE} observatory
consists of four coaligned prime-focus telescopes (two SiC and two
LiF) and Rowland-circle spectrographs. The SiC gratings provide
reflectivity over the range 905-1105 \AA, while the LiF have
sensitivity in the 990-1187 \AA~range. Each detector is composed of
two micro-channel plates, therefore a gap of $\approx$ 5 \AA~divides
each of our spectra into two pieces.  The list of the four targets
studied in this work and the log of the observations are presented in
Tables~\ref{targets} and~\ref{log} respectively. All data were
obtained with the source centered on the 30'' $\times$ 30'' (LWRS)
aperture with total exposure times ranging from 4.8 ks (HD\,192639) to
25.5 ks (HD\,73882).  All our datasets have a S/N ratio per pixel
around 10.  The data were processed with version 2.0.4 of the CalFUSE
pipeline.  Corrections for detector background, Doppler shift,
geometrical distortion, astigmatism, dead pixels, and
walk\footnote{ The FUSE detector electronics happens to miscalculate
the X location of photon events with low pulse heights. This effect is
called ``walk''.} were applied, but no correction was made for the
fixed-pattern noise. The 1-D spectrum was extracted from the 2-D
spectrum using optimal 
extraction\footnote{http://fuse.pha.jhu.edu/analysis/lacour/}
\citep{1986PASP...98..609H,1986PASP...98.1220R}. Instead of co-adding
the different segments of the spectrum, we used only the segments that
appear to have the best correction of the distortion effects. Those
are SiC2A (930-990 \AA), LiF1A (990-1080 \AA), SiC1A (1080-1088 \AA),
and LiF2A (1090-1187 \AA).  Below 930 \AA, the high reddening of our
targets, removing most of the flux, does not allow for reliable
measurements.

After binning the data by 4 pixels ($\approx$ 7 km\,s$^{-1}$), the
processed data have a S/N ratio of nearly 20 per bin, and a nominal
spectral resolution of $\approx$ 20 km\,s$^{-1}$ (FWHM).

\section{H$_2$ Measurements}
\label{sec:targe}

The {\it FUSE} wavelength range allows us to access a large number of
H$_2$ absorption lines, corresponding to a wide range of rotational
excitations. For each of the levels that we focus on ($J = 2-7$), we
have measured, when available, column densities and $b$ values. To
ensure consistency of the measurements, we used two
different methods to determine $N$ and $b$, described below.

\subsection{Curve of Growth Method (COG)}

The measured equivalent widths (EqW) of each line studied in this
work are summarized in Table~\ref{EqW}. The stellar continuum in the
vicinity of each line was estimated using a low-order Legendre
polynomial fit to the data. The 1$\sigma$ error bars were computed
taking into account four types of errors, added in quadrature:

\begin{itemize}
\item The statistical errors, supposed to be a white Poissonian
noise. These errors (roughly the square root of the count rate) are
computed by the pipeline for each pixel. The total statistical error
over each line is therefore the quadratic sum of the error of each
integrated pixel \citep[more information can be found in Appendix A
of][]{1992ApJS...83..147S}.
\item The background uncertainties, proportional to the exposure
time. They have been estimated by the {\it FUSE} science data
processing team to be at the level of 10 $\%$ of the computed
background\footnote{http://fuse.pha.jhu.edu/analysis/calfuse\_wp3.html}.
This error is calculated by the pipeline, and added to the statistical
errors.
\item The continuum placement error, which depends mainly on the S/N
ratio in the vicinity of the line. To estimate this error, we shift
the continuum by 1 to 3 $\%$ (depending on the S/N ratio), calculating
a lower and upper value for the EqW. The difference is taken as
the 1$\sigma$ error. We note here that because of to the
stellar type of the targets (see $Sp.T.$ in Table~\ref{targets}), very few
stellar lines are present, and are easily distinguishable with the
interstellar line due to their thermal broadening.
\item The systematic uncertainties, which are the most difficult
errors to quantify. They may come from geometrical distortions, walk,
dead pixels, point spread function (PSF), fixed pattern noise, etc.
Most of these distortions are corrected by the pipeline, but these
effects may nevertheless have a non-negligible influence on our
measurements.  Moreover, there is no way to estimate the effect over a
single absorption line.  Assuming that systematic errors are
homogeneous over our measurements, we adjusted the systematic errors
to be proportional to the EqW. The factor of proportionality is set so
that the total $\chi^2$ of the COG fit is equal to the number of
degrees of freedom. To avoid any bias, a proportionality factor was
obtained independently for each sight line and each species (i.e., for
each $J$ level), which is possible because the number of spectral
lines being measured is statistically significant.  The resulting
factors are in the range of 1 to 8 $\%$.
\end{itemize}

To determine column densities and $b$ values, we fit a set of 300
single-Gaussian curves of growth to our measured EqWs. They were
obtained by integrating a Voigt profile over a large number of $b$
values and damping factors. For each species, $\chi^2$ was calculated
for each $b$ value and column density ($N$).

\begin{figure}
\centering
\includegraphics[angle=90,height=19cm]{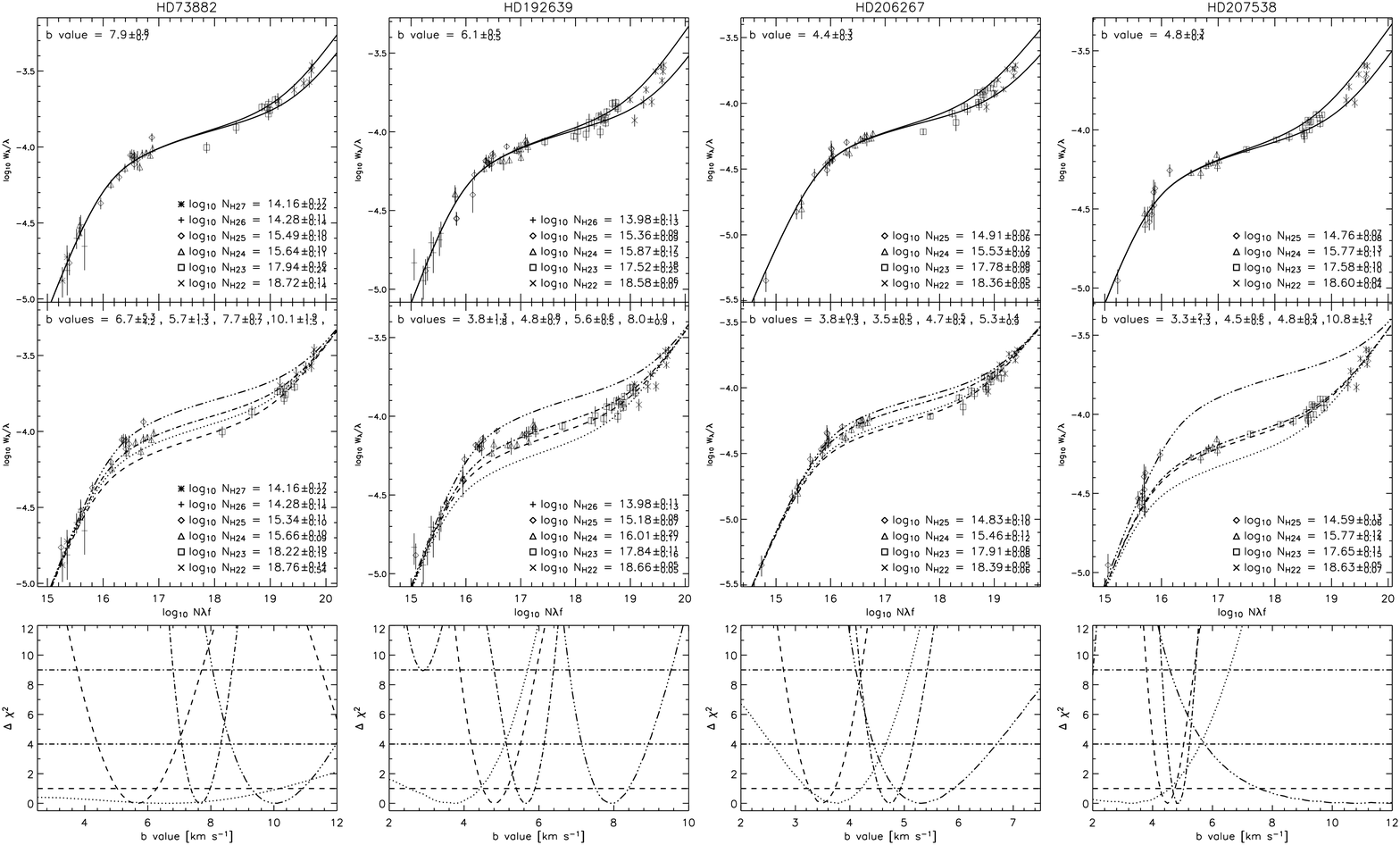}
\caption{Curve of growth analysis for H$_2$ along the HD\,73882,
HD\,192639, HD\,206267 and HD\,207538 lines of sight. The plots at the
top assume a common $b$-value for all $J$ levels. In the middle plots,
each $J$ level is fit with a different $b$-value. The plots in the
bottom of the figure present $\Delta\chi^2$ as a function of $b$ for
the COGs displayed in the middle of the figure.  In the $\Delta\chi^2$
plots, the different rotational levels are represented with dotted
($J$ = 2), dashed ($J$ = 3), dot-dashed ($J$ = 4), and
dot-dot-dot-dashed ($J$ = 5) lines. Note that due to the non-linearity
of the COG, the $\Delta\chi^2$ sometimes shows two minima.  In these
cases, the second minimum is ruled out on the grounds that the
resulting column densities are not physical.
\label{COGs}}
\end{figure}

The best COG fits are shown in Figure~\ref{COGs}.  The upper plots
show the resulting COG using a single $b$ value for all the rotational
levels. The value of $\chi^2$ compared to the number of degrees of
freedom is the best mathematical tool to evaluate the goodness of a
fit. For these fits, the values are 52/35 for HD\,73882, 110/58 for
HD\,192639, 54/44 for HD\,206267, and 37/36 for HD\,207538. The
probabilities of having $\chi^2_{\nu}$ equal to or larger than those values
are 3$\%$ ($> 2 \sigma$), 0.0045$\%$ ($> 4 \sigma$), 14$\%$ ($> 1
\sigma$) and 42$\%$ respectively.  Therefore, the result is not
significant for HD\,207538 but HD\,192639, and to a lesser extent,
HD\,73882 and HD\,206267, clearly have an inconsistency in the fit.
The middle plots display our best fits using different $b$ values for each
rotational level and the bottom plots show the $\Delta \chi^2$ as a
function of $b$, again, for each rotational level. All errors listed
here are at the 2$\sigma$ level, corresponding to a $\Delta \chi^2$ of
4.

To check the possibility that some systematic problem in one (or a
few) lines could induce a ``false broadening effect'' on the
HD\,192639 dataset, we randomly removed half of our measurements, then
the other half. The results, summarized in Table~\ref{COGtest},
confirm the presence of the broadening.

\subsection{Profile Fitting Method \label{PFMethod}}

\begin{figure}
\epsscale{0.6}
\plotone{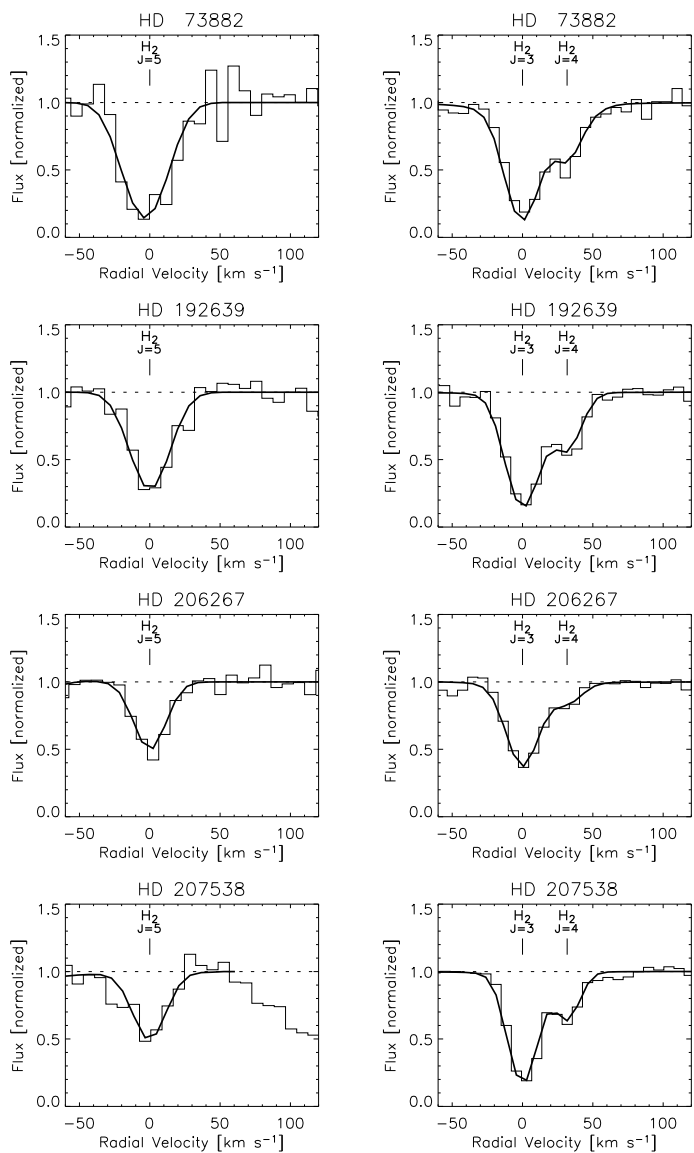}
\caption{Best profile fits to the $J=5$ 1017.8 \AA~line (left) and
$J=3$ 1115.9 \AA~and $J=4$ 1116.0 \AA~lines (right), for the four
sightlines. Each rotational level is fit with a different $b$
value. The continuum is normalized to 1.\label{PFBdiff}}
\end{figure}

\begin{figure}
\epsscale{0.75}
\plotone{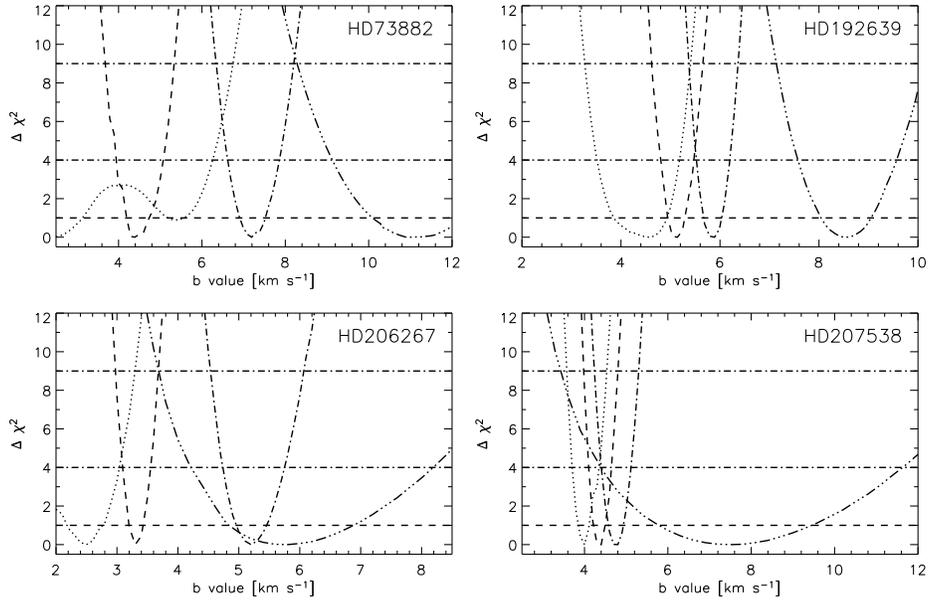}
\caption{$\Delta \chi^2$ curves for the different $b$ values used with
profile fitting.  The different rotational levels are represented with
 dotted ($J$ = 2), dashed ($J$ = 3), dot-dashed ($J$ = 4), and
dot-dot-dot-dashed ($J$ = 5) lines.
\label{PFxi2}}
\end{figure}

The spectral resolution ($\approx$ 20 km\,s$^{-1}$, varying over
several km\,s$^{-1}$, depending on the detector used), is insufficient
to directly determine variations in the broadening of lines, which are
typically of the order of a few km\,s$^{-1}$. Instead, it is the
relative shape of the lines, combined with knowledge of the oscillator
strengths, that provides meaningful information. We used a profile
fitting routine, {\it Owens}
\citep{2002ApJS..140...67L,2002ApJS..140..103H}, developed by
M. Lemoine at the Institut d'Astrophysique de Paris, which allows
fitting all the lines simultaneously while considering different $b$
values for each rotational level. To minimize systematic errors
induced by the PSF, we allowed the width of the PSF to vary by
$\approx$ 25 $\%$ about the nominal value (15-25 km\,s$^{-1}$). An
important advantage of fitting multiple species is the added ability
to work with semi-blended lines. The lines that were included with
this method are listed in Table~\ref{EqW}.  This was particularly
helpful in constraining the $J$ = 2 $b$ value, because of the weak
oscillator strength of the 1112.5 \AA~blended
line. Figure~\ref{PFBdiff} shows some sample best fits for the 1017.8
\AA~($J$ = 5), 1116.0 \AA~($J$ = 4) and 1115.91 \AA~($J$ = 3)
lines. In Figure~\ref{PFxi2} we plot the $\Delta \chi^2$ for the $b$
values used in the fits.  To account for systematic uncertainties, we
scaled the errors (by a factor of 1 to 2) so that the $\chi^2$ minimum
is equal to the number of degrees of freedom \citep[for more on the
$\chi^2$ technique while using {\it Owens},
see][]{2002ApJS..140..103H}.

\begin{figure}
\epsscale{0.6}
\plotone{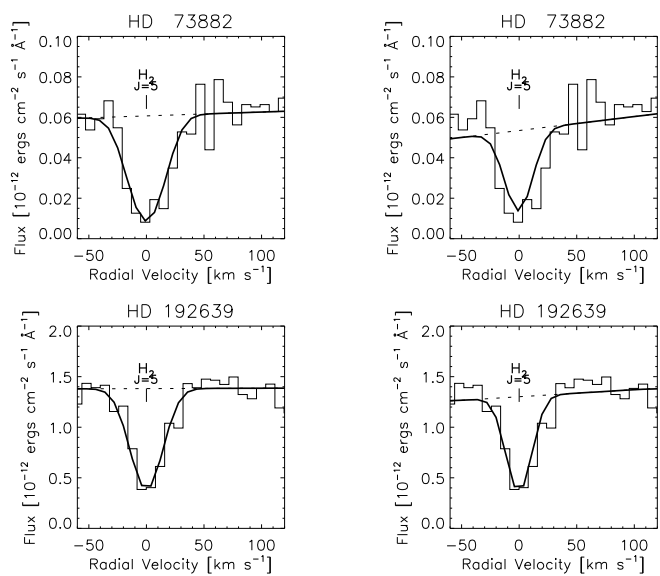}
\caption{Best profile fits of the $J=5$ 1017.8 \AA~line for HD\, 73882
(top) and HD\, 192639 (bottom) using a single $b$ value for all $J$
levels (right) and allowing different $b$ values for each $J$ level
(left). The inadequacy of the single $b$ value fits for all $J$ levels
is apparent both from close inspection of the fits and from the
reduced $\chi^2$ values (see section~\ref{PFMethod}).
\label{PFBsingle}}
\end{figure}

To convince ourselves that profile fitting with a single $b$ value for
all levels does not accurately describe our data, we did the fitting
for HD\,73882 and HD\,192639 and compared the fitting on one
particular line.  The 2 panels in the right column of
Figure~\ref{PFBsingle} show the best fit for the 1017.8 \AA~line
(segment LiF1A) when a single $b$ value was used for all rotational levels,
while the left panels show the best fits using individual $b$
values. Without taking systematic uncertainties into account (i.e.,
without scaling the errors), the $\chi^2$ for the left panels are
close to the number of degrees of freedom (28/27 for HD\,73882 and
24/21 for HD\,192639), indicating that the fits are reliable. On the other
hand the $\chi^2$ for the right panels (62/27 and 44/21 respectively)
indicate an inconsistency. Note that the PSF was a free parameter and
had a value of 2.96 binned pixels (23.4 km\,s$^{-1}$) for HD\,73882
and 2.98 binned pixels (23.2 km\,s$^{-1}$) for HD\,192639 (left
column). For the right column, the obtained PSFs are 2.67 pixels (21.1
km\,s$^{-1}$) for HD\,73882 and 2.42 pixels (19.1 km\,s$^{-1}$) for
HD\,192639. The differences in the PSFs are within the {\it FUSE}
resolution uncertainties \citep[see][]{2000ApJ...538L...1M}.

\subsection{Results}

Tables~\ref{tb:res} list the column densities and the $b$ values,
respectively, derived using the methods described above (2$\sigma$
uncertainties). Profile fitting allows us to quickly determine upper
limits on column densities for $J$ levels which only have transitions
that are too weak to be used with the COG method. Column densities for
the $J$ = 0 and 1 levels, for the four stars, are from
\citet{2002ApJ...577..221R}. The $b$ values are consistent between PF
and COG. In each case, but with different reliability levels, the
velocity dispersion shows the same increasing trend with the H$_2$
excitation levels. Since $b$ values and column densities are
interdependent, it is an important result for observers investigating
saturated H$_2$ lines. If the broadening of the higher $J$ levels is
assumed to be equal to those of the lower $J$ levels, then there is
the possibility of considerably overestimating the column densities.

\section{On the source of excitation and broadening}
\label{sec:conc}

\begin{figure}
\epsscale{.9}
\plotone{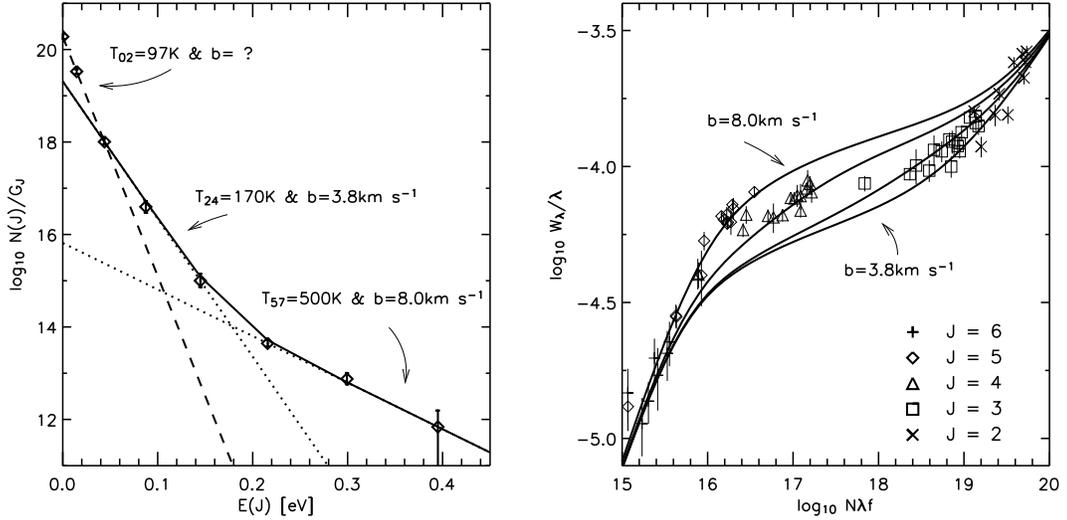}
\caption{{\it Left panel :} Excitation diagram of the H$_2$ along the
 HD\,192639 line of sight. We fitted the column densities with several
 components, with different excitation temperatures and velocity
 dispersions. The $T_{ex}=T_{J=2-4}$ component is supposed to have a
 broadening equal to the $J = 2$ species, and the $T_{ex}=T_{J=5-7}$
 component a broadening equal to the $J = 5$ species. {\it Right
 panel :} Two components curve of growth for $J = 2$ to 6, with a
 relative strength in accordance with the two components fitted on the
 higher levels of the excitation plot. A variation of the $b$-values
 appear due to the presence of the two components.
\label{multiCOG}}
\end{figure}

In light of the results presented above a question must then be asked:
What causes the increase in the broadening of the H$_2$ lines with
increasing $J$ level?  An obvious explanation would be that we may
have a broad component revealed at high rotational level by an high
excitation temperature. To test this possibility, we plot in
Figure~\ref{multiCOG} the curves of growth for a sightline having two
components with different velocity dispersions. Because the relative
strength differs from one rotational level to the other, each level
correspond to a curve of growth with a different shape. As an example,
we used the excitation diagram of target HD\,193639 to fit two
components which are associated with the $J = 2$ level for one, and
the $J = 5$ level for the other. We then fitted the EqWs on the COGs
of their corresponding excitation level. The fit explain why an effect
due to the variation in the ratio between two components, one broad and
the other narrower, can be seen as a variation in the broadening. We
note nevertheless that this explanation is incompatible with Component
1 observed with {\it IMAPS} towards $\zeta$ Ori A
\citep{1997ApJ...477..265J}. In this case, the third rotational level
column density is so low, that it cannot belong to a different
component.

Finding an explanation for the presence of a broad component is
another challenge. It may be the key behind the source of both the
excitation of H$_2$ and the presence of large amounts of
CH$^+$. Specifically, the fast ion-molecule reaction, CH$^+$ + H$_2$
$\longrightarrow$ CH$_2^+$ + H, in cold gas, predicts CH$^+$ column
densities far below the observed levels
\citep{1974ApJ...189..221W}. This is also the case toward our
sightlines ($N({\rm CH^+}) > 10^{13}$ cm$^{-2}$).  The solution might
be in warm interstellar gas ($T \geq 10^3$ K) in which the endothermic
reaction C$^+$ + H$_2$ $\longrightarrow$ CH$^+$+ H - 0.4eV can provide
an equilibrium density close to the observed levels. The presence of a
warm component received strong support by the observation of a
correlation between CH$^+$ and the rotationally excited H$_2$
\citep{1986ApJ...303..401L}. However, since until now no direct
observations of this warm component have been obtained, parameters
such as its density and temperature are unknown. The increase of $b$
with increasing $J$ level seems to be direct evidence of a warm
component. Several excitation mechanisms, discussed below, could be
responsible for this effect.

\subsection{UV-pumping}

The ro-vibrational cascading releases its energy through infrared
photons \citep{1976ApJ...203..132B}. Like photoexcitation, such energy
loss does not change the kinetic energy of the molecules, and
therefore it does not affect the velocity dispersion. Heating of the
gas can nevertheless occur, through photodissociation of H$_2$ and
photoelectron emission from dust grains. However, such process would
require a high UV field \citep[as the one towards the Pleiades cluster,
e.g.][]{1984ApJ...284..695W}, and is an unlikely explanation for a
broadening of up to 10 km\,s$^{-1}$.

\subsection{H$_2$-formation pumping}

When molecules are created on the surface of grains, they carry away
most of the initial energy ($\approx$ 4.5 eV) which provides the
kinetic, rotational and vibrational excitation of H$_2$.  Support for
this mechanism was obtained by \citet{1992MNRAS.259..155W} who
calculated the column density ratio between the $J$ = 4 to 7 levels in
good agreement with observations. However, absolute column density
calculations do not exist. To address this, we constructed a
time-dependent model in which we followed the stochastic evolution of
the molecules after their grain formation. Details of the model are
given in the Appendix. According to this model, dispersions appear as a
natural consequence of the equilibrium among the various excitation
and decay processes. Figure~\ref{Broadeps} shows the expected
broadening, as a function of the density and the rotational
level. Using the expected broadening in conjunction with
Equation~\ref{eq:dens} we are able to calculate the densities and
formation rates needed to explain both the velocity dispersions and
the column densities of the broader rotational levels ($J \geq
4$). The densities are in agreement with previous analysis of the
\ion{C}{1} fine-structure excitation for HD\,192639 and HD\,206267,
which led to to estimated densities of 16 cm$^{-3}$
\citep{2002ApJ...576..241S} and 30 cm$^{-3}$
\citep{2001ApJS..137..297J}, respectively.  However, the formation
rates implied by our models ($R$ in Table~\ref{tb:resmodel}), do not
agree with previously calculated values \citep[$R \approx 3 \times
10^{-17}$ cm$^3$\,s$^{-1}$
in][]{1975ApJ...197..575J,2002A&A...391..675G}. There is a factor of
approximately ten between the rate needed to explain the column
densities and the rates mentioned above.

A second argument against H$_2$-formation pumping as responsible for
the broadening of the excited states comes from the difficulty to
account for the CH$^+$ column density. We report on the upper right
panel of Figure~\ref{modeleps} the threshold energy for the
C$^+$+H$_2$ reaction, and found that the time during which the
molecule is kinetically warm is $t_{\rm warm} = 4 \times 10^{9}/n$ s
(the cooling time is roughly inversely proportional to the density $n$, in
cm$^{-3}$). Assuming that the density is spatially uniform along the
sightline, we can obtain the column density of warm H$_2$ as a
function of the atomic hydrogen and the formation rate : $N({\rm
H_2})_{\rm warm} = N({\rm H})\,nR\,t_{\rm warm}$. Considering the
formation rates from Table~\ref{tb:resmodel}, we obtain a ratio
$N({\rm H_2})_{\rm warm} / N({\rm H}) \approx 1.2 \times 10^{-7}$, a
value several orders of magnitude below what is needed to explain the
column density of CH$^+$ \citep[$N({\rm H_2})_{\rm warm} / N({\rm
H_2})_{\rm cool} \approx 10^{-3}$ in][]{1986ApJ...303..401L}.

\subsection{Collisional excitation in a warm environment}

Warm low density interstellar gas surely is present along the lines of
sight.  \citet{1969ApJ...155L.149F} were the first to show that warm
gas could be thermally stable at low densities.  Such gas ($n_{\rm H}
\approx 0.1$ cm$^{-3}$; $T \approx 7000$ K) appears to be the major
constituent -- in mass -- of the local interstellar cloud
\citep{1983ApJ...271L..59F,1996SSRv...78..157L,1998LNP...506...19L},
but a recent survey of the LISM, performed by
\citet{2003ApJ...595..858L}, showed that in this medium, the H$_2$
molecular fraction is low, close to 10$^{-5}$.  The problem appears to
be that at such densities and temperatures, H$_2$ formation on grains
becomes negligible. Other routes exist, such as formation through
hydrogen ions \citep{1981ApJ...249..138B}, but the rates for the
process are low, as is the observed amount of such ions
\citep{2002AAS...201.4709A}. The same arguments can be used against
the presence of H$_2$ in the warm postshock gas of dissociative
shocks. However, \citet{1997ApJ...477..265J} concluded anyway that in
the line of sight towards $\zeta$ Ori A, the dispersion of the $J = 5$ line
could be the trace of an ongoing J-type shock after which the gas
recombines and cools. It is true that this theory was corroborated by
a shift of the line centers - the higher $J$ levels being shifted
towards lower velocities - which is difficult to explain otherwise.

But shifts could also be the trace of slower shocks, C-type
shocks. Even though the conditions in the postshock gas are not
favorable in terms of H$_2$ formation, the fact that it is
non-dissociative makes it possible to contain a large amount of heated
H$_2$. Hence, \citet{1978ApJ...222L.141E} showed that such shocks
could heat a large portion of the cloud with temperatures of several
thousands Kelvin. This mechanism would generate both the broadening
and the CH$^+$ column densities.  However, there are several arguments
against the presence of a single important shock front: predictions of
$N$(OH) produced via an endothermic reaction would be significantly
higher than what is observed, and significant velocity shifts required
for this type of shocks are usually undetected. We can however derive
the environment parameters in the hypothesis of a thermal broadening
of the postshock gas. We listed in Table~\ref{tb:warm} the kinetic
temperature (under the assumption that the linewidths are mostly due
to thermal broadening), and the excitation temperature (derived from
the $J = 5$ and 7 levels) for the four lines of sight. The difference
between the two temperatures constrain the ratio between collisional
excitation and radiative de-excitation rates. Hence, we used the rates
from \citet{1999MNRAS.305..802L} and \citet{1998ApJS..115..293W} to
infer the densities, and therefore the pressure $log(P/k) = log(n
T)$. We note that we are not able to derive lower limits of the $J =
7$ rotational column density. However, we used the best fitting values
to infer the pressure of the neutral gas, and obtained values
significantly higher than those derived from the \ion{C}{1} fine
structure; log(P/k) $\approx 3.1$ for HD\,192639
\citep{2002ApJ...576..241S} and 3.5 for HD\,206267
\citep{2001ApJS..137..297J}.

From the pressure calculations, we argue that we are more likely in
the presence of material cooler than what can be derived from a
thermal velocity dispersion. The molecule must still be hot enough to
explain the excitation temperature (approximatively a thousand
Kelvin), but not up to the temperature required for a thermal
broadening ($T^{k}$ in Table~\ref{tb:warm}). Hence, the only remaining
explanation of the velocity dispersion is that we are looking at a
warm {\em and} turbulent layer, most probably intimately associated
with the cold medium. To explain the presence of such layer, some
invoke supra-thermal velocities of ions relative to the neutrals
driven by multiple magneto-hydrodynamic (MHD) criss-crossing shocks
\citep{2002A&A...389..993G}.  Others invoke the intermittency of
turbulence and the existence of localized tiny warm regions,
transiently heated by bursts of ion-neutral friction and viscous
dissipation in coherent and intense small small vortices
\citep{1998A&A...340..241J}. The main physical difference between each
phenomenon is the thickness and the crossing time.  While a warm MHD
postshocks layer can have a thickness of $\approx 0.1$ pc, coherent
vortices threaded by magnetic fields may have radii as small as
$\approx 20$ AU. Both would achieve peak temperature around 1000K, but
with differential velocities around 10 km\,s$^{-1}$ for the MHD shocks
and around 4 km\,s$^{-1}$ in a coherent vortice.

Towards our targets, the situation would be perfectly describe by both
explanations. The many small-scale shocks or vortices would create the
observed velocity dispersion, while embedded magnetic fields would
generate differential velocities between ion and neutral species,
reacting into CH$^+$ through the C$^+$ + H$_2$ $\longrightarrow$
CH$^+$+ H endothermic reaction. It would also give an answer to
multiple quests for a warm layer \citep[e.g. CO, HCO+, H$_2$O
in][]{2000A&A...356..279P,2000A&A...355..333L,2002ApJ...580..278N}.
However, other heating process may be possible, and are difficult to
rule out due to our low spectral resolution. The probability
distribution functions of these high rotational levels --- either in
FUV or IR \citep[Falgarone et al. in press]{1999usis.conf..779V} ---
would eventually help us to distinguish between one process and the
other.

\section{Summary}

We observed four highly reddened ($E_{(B-V)} \approx 0.6$) lines of
sight in which the saturation of the higher rotational levels of H$_2$
allows us to infer their velocity dispersion. We measured broadenings
up to 10 km\,s$^{-1}$, increasing with the energy of the rotational
level. Considering the fact that it was already observed towards
several other sightlines
\citep{1973ApJ...186L..23S,1974ApJS...28..373S,1996AAS...188.0706J},
we suggest that this phenomenon is a fairly common one. As a first
result, we suggest caution in investigating saturated H$_2$ lines
since assuming an identical linewidth for all the rotational levels
could lead to significant systematic errors on the column densities.

We looked at the possible sources of rotational excitation to see
which could induce such broadening of the absorption lines. We ruled
out UV-pumping which can not create a velocity dispersion by
itself. We constructed a time dependent model of the state of the
molecule following formation on the grain. From it, we deduced that
such mechanism may be responsible for the broadening, but would however
need a formation rate ten times the one derived from previous
studies \citep{1975ApJ...197..575J}. We also note that the amount of
warm H$_2$ created would not be enough to account for the observed
column densities of CH$^+$.

We conclude that the more likely explanation is the presence of a
turbulent, warm layer in the molecular cloud. The temperature would
need to be over 600 K to account for the rotationally excited H$_2$,
and the velocity dispersion of the gas patches around 8
km\,s$^{-1}$. Small criss-crossing shocks or vortices could be the
phenomenon behind this turbulent layer.  Magneto-hydrodynamic waves
created inside them would convert the kinetic energy to create the
observed amount of CH$^+$ through its endothermic reaction with H$_2$.



\acknowledgments

The authors thank N. Balakrishnan for his work on the modeling of the
cooling of hot H$_2$. The authors also thank Ed Jenkins for invaluable
discussions, thorough comments on an early draft and analysis of
\ion{Cl}{1} lines which do not figure in this paper. Further thanks go
to E. Falgarone, B. Racheford, P. Sonnentrucker and N. Lehner for
helpful comments and discussions.  SL would also like to personally
thank the entire JHU {\it FUSE} Team for their hospitality, their
knowledge and their kindness in sharing it. Part of this work has been
done using the profile fitting procedure {\it Owens.f} developed by
M. Lemoine.  French participants are supported by CNES.  This work is
based on data obtained for the Guaranteed Time Team by the
NASA-CNES-CSA {\it FUSE} mission operated by The Johns Hopkins
University.

\begin{appendix}

\section{Modeling  H$_2$ formation}

\subsection{Description}

In this section we describe our numerical model of H$_2$ excitation
and de-excitation following H$_2$ formation on grains. We use it to
show the existence of a possible correlation between the excitation
level of the molecule and the line broadening associated with that
level. We also use it to calculate the amount of excited H$_2$
obtained from H$_2$-formation pumping.  Note that throughout most of
this section we are dealing only with the gas that is excited by
formation pumping. The effects described here are not affected by the
presence of another gas component (at the same level) excited by a
different mechanism.

The formation of H$_2$ on grains has been discussed at length in the
literature
\citep[e.g.][]{1963ApJ...138..393G,1966BAN....18..256K,1970ApJ...162..463A,
1971ApJ...163..155H,1972NPhS..237...99L}.  Our model is based on the
fact that when H$_2$ molecules are formed, they will leave the surface
of the grain carrying with them excess energy, which will be
distributed as surface, kinetic, rotational and vibrational energy.
If we can determine the cross-sections for the various processes by
which the molecules either lose or convert energy from one form into
another, we can then estimate the average kinetic energy over the
lifetimes of each species. Adding in the formation rates, we are able
to predict the column density and velocity dispersion of each level.

\subsubsection{Initial distribution}

Recent simulations of H$_2$ formation have been performed for graphite
surfaces \citep{1998A&A...334..363P,2001..Meijer} and icy interstellar
dust \citep{1999ApJ...520..724T}. In all cases, the molecules leave
the grain with an initial energy of 4.48 eV, but the distribution among
binding, kinetic, and rotational energy differ considerably.
We based our model on the quasi-classical computer simulation of the
Parneix \& Brechignac paper, due to their choice of computational
method and physical model, which includes an entrance-channel barrier
on the potential energy surface of the grain. The initial distribution
over the vibrational levels corresponds to Fig. 11 of their paper, and
averages around 1.1 eV, with a peak at $v$ = 0. The distribution of
rotational energy, with a mean value of 0.7 eV, corresponds to the
Fig. 9 from the same paper, and is the same for each vibrational
level. 1.7 eV is carried as kinetic energy, and the remainder is used
to desorb the molecule from the grain.  The implications of these
choices are discussed in the ``Model results \& considerations''
Section. The ortho/para repartition (OPR) is an important factor in the
column density calculations, but fortunately, it does not affect the
results concerning the velocity dispersions. In view of the results of
\citet{1995..Persson}, we decided not to include the statistical
weight factor of 1/3.

\subsubsection{Kinetic cooling}

Starting with an initial kinetic energy, the H$_2$ molecule will cool
down, while interacting with its environment. Because the formation
rate is proportional to the density of atomic hydrogen, most of the
formation of molecular hydrogen takes place in the photodissociation
region, where hydrogen is mostly atomic. Moreover, simulations
\citep{2002A&A...390..369L} show a sharp decrease of the molecular
fraction as we go deeper in the cloud, implying a formation rate only
marginal inside the cloud. It leaded us to approximate the formation
environment as being purely atomic.  Under this assumption, we
needed just three parameters to compute the kinetic cooling: the
H-H$_2$ cross-section ($\sigma$), the mean energy loss per collision
($\gamma$), and the density ($n$). While $n$ is just a property of the
medium, the two other parameters depend on the energy of the molecule.
To obtain $\sigma$, we extrapolated to higher energies the values in
Fig. 2 of \citet{1995..Clark}. The results are plotted in the upper
left panel of Figure~\ref{modeleps}.  $\gamma$ was generated using
equations (12) and (25) in \citet{1998JASTP..60..95}, while $n$ is
left as a free parameter, varying from 0.1 to 1000 cm$^{-3}$. The
right upper panel of Figure~\ref{modeleps} displays the estimated
kinetic energy as a function of time for a medium with density $n =
10$ cm$^{-3}$. One can see that the translational energy is lost
fairly slowly, allowing time for the molecule to undergo endothermic
reactions.  For example, if we consider the threshold energy for the
C$^+$ + H$_2$ $\longrightarrow$ CH$^+$+ H - 0.4eV reaction (dashed
line), the molecule would have such or more energy during $t_{warm} =
4 \times 10^8$ seconds ($\approx 10$ ans).

\subsubsection{Radiative cooling}

From the initial distributions, we generated a table of 20 rows
(corresponding to the 20 first rotational levels) and 7 columns
(corresponding to the 7 first vibrational levels). In steps of 10$^5$
seconds, and using the radiative decay table from
\citet{1998ApJS..115..293W}, we computed the density of each
excitation level from its formation until 3.5 $\times$ 10$^{11}$
seconds. The panel in the center of Figure~\ref{modeleps} shows the densities
(normalized to 1) of each vibrational level (independent of the
rotational level). After a fairly short time ($\approx 10^7$s), all of
the molecular hydrogen is in the ground vibrational state. Then, the
model clearly shows the progression towards lower $J$ levels as time
increases (lower panels of Figure~\ref{modeleps}). The rising parts of
the density curves are caused by the cascading down from higher $J$ or
$v$ levels, while the steep drops occur at the radiative lifetime.
Interestingly, these lifetimes (e.g., A$_{42}^{-1} \approx 10^9$ s)
are of the order of the kinetic cooling time (see previous paragraph). It
follows that, depending on the density of the cloud, some rotational
levels will be kinetically hot, while lower levels will not.

\subsubsection{Inelastic collisional cooling \& excitation}

In addition to the decay rates, collisional excitation and
de-excitation may have an important effect, especially while the
molecule is still highly energetic, or when the density is high. We
used the rates from the web site {\it
http://ccp7.dur.ac.uk/cooling\_by\_h2/index.html}
\citep{1997ApJ...489.1000F,1999MNRAS.305..802L}, in addition to the
decay rates of the above section, to model the time dependent density
distribution. The lower panels of Figure~\ref{modeleps} show the resulting
densities (normalized to 1) of the pure rotational levels. The
specific conditions for this plot are a density of 10 particles per
cm$^3$, and an ambient temperature of 100 K. As discussed in the 
previous paragraph, it is apparent that the fraction of each
excited state falls off rapidly at a time corresponding to the decay
rate, and that inelastic collisions are not an important de-excitation
mechanism.  Numerical tests show that for densities below several
hundred cm$^{-3}$, inelastic collisions are negligible compared to
radiative decay.

\subsection{Model results \& considerations}

\begin{figure}
\epsscale{.75}
\plotone{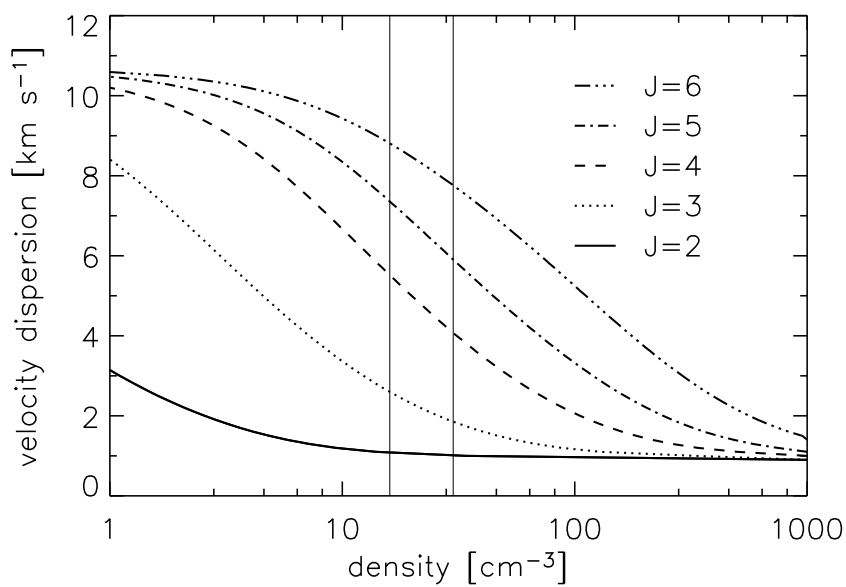}
\caption{Line broadening ($b$) derived from the mean kinetic energy of
each level following formation on a grain. Since the value of the 
kinetic energy is dominated by collisional cooling, the broadening is
a strong function of the density ($n$). The thin vertical lines show
the density towards HD\,192639 \citep{2002ApJ...576..241S} and HD\,206267
\citep{2001ApJS..137..297J}.
\label{Broadeps}}
\end{figure}

\begin{figure}
\epsscale{.56}
\plotone{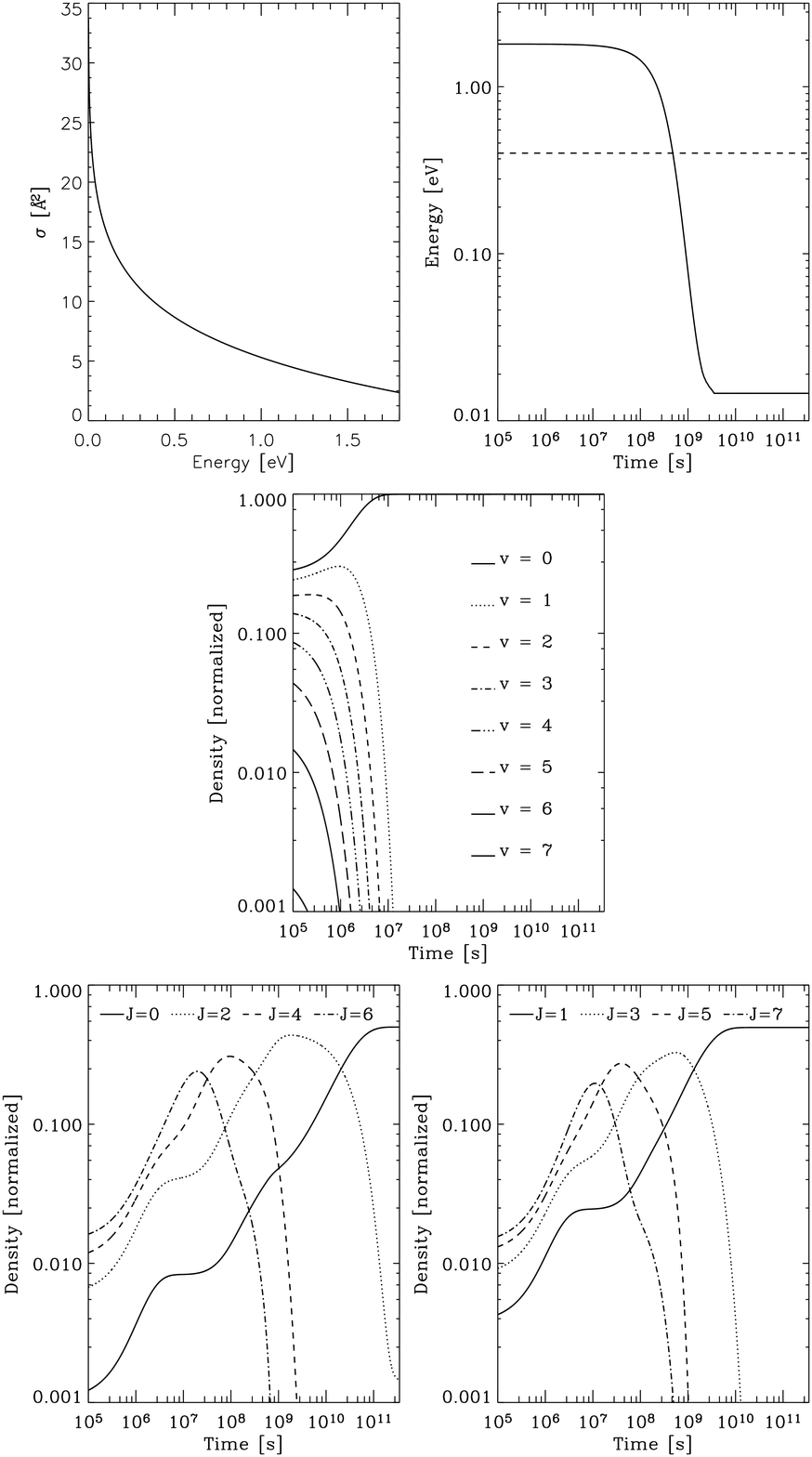}
\caption{Time dependent model of molecular hydrogen following
formation on grains (see Appendix). {\it Upper left panel :} Effective
H-H$_2$ cross-section as a function of the kinetic energy. {\it Upper right
panel :} Kinetic energy of the H$_2$ molecule as a function of
time. The dashed line corresponds to the energy threshold from which
C$^+$ can react with H$_2$ to form CH$^+$. {\it Three lower panels :}
Distribution of H$_2$ molecules over the vibrational and the ortho and
para rotational states, as a function of the molecule's lifetime. Note
that radiative vibrational transitions ensure decay towards the
vibrational ground state on short time scales. The initial rise of the
rotational density curves is due to the cascading of molecules down
from the higher $J$ and $V$ levels, whereas the steep drops occur at
the radiative lifetime of the level.  Because collisional excitation
and de-excitation are part of our simulation, all these plots depend
on the density, here equal to 10 cm$^{-3}$.
\label{modeleps}}           
\end{figure}

Integrating the normalized column densities over time allows us to
 extract two pieces of information :
\begin{enumerate}
\item The mean kinetic energy of each rotational levels, which can be directly
linked to the velocity dispersion as long as we assume that the
excitation is mainly due to its formation on the grain. The mean
energy is obtained by:
\begin{equation}
E_k(J,n)=\frac{\int^{+\infty}_0\mathrm{Density}(J,n,t)\times
E(t)dt}{\int^{+\infty}_0\mathrm{Density}(J,n,t)dt}
\end{equation}
In practice the upper limit of the integration was set to 3.5 $\times
10^{11}$ seconds, by which time the excited species have largely
decayed.  The velocity dispersion, i.e., the broadening, is therefore
\begin{equation}
b=\sqrt{4/3*E_k/(2m_{\rm H})}\,. 
\label{eq:b}
\end{equation}
We plotted in Figure~\ref{Broadeps}, the broadening of each rotational
levels between 2 and 6, clearly showing that it is increasing with the
rotational level. Again, this is only the case when the species are
populated by H$_2$-formation pumping.  The broadening of higher
rotational levels ($J = 4$ and 5) were fitted on the curves, and
estimated densities are reported for our four targets in
Table~\ref{tb:resmodel}.
\item Still assuming no other source of excited H$_2$, we are also able to
calculate the molecular hydrogen column densities of the excited
levels.  The densities depend on the
formation rate ($nR$), and the atomic hydrogen density. Assuming that
the density is spatially uniform along the line of sight, we can use
the column density of atomic hydrogen $N(\rm H)$, and therefore,
have the following equation:
\begin{equation} 
N({\rm H_2})(J,n) = N({\rm H})\,nR\,\int^{+\infty}_0\mathrm{Density}(J,n,t)dt 
\label{eq:dens}
\end{equation}
To remove the effect of the formation OPR on the column
density, we considered the sum of the column densities of the $J = 4$
and 5 levels. The formation rates needed to create such column densities
are reported in Table~\ref{tb:resmodel}.
\end{enumerate}
The distribution of the formation-energy of H$_2$ has been calculated
through many other theoretical models, yielding quite different
results in all the different parameters. Two parameters have a large
influence on our results: the initial kinetic and rotational
energies. On one hand, the kinetic energy will decide the amplitude of
the velocity dispersion. For instance, whereas our model starts with a
kinetic energy of 1.7 eV, \citet{2001..Meijer} find an initial energy
of 1.18 eV which would scale down all our broadening calculations by a
factor of 1.2. More specifically, from Equation~\ref{eq:b}, a
broadening of 8 km\,s$^{-1}$ can be explained if the kinetic initial
energy is equal or above 0.95 eV. This is compatible with most but a
few of the theoretical formation models.  On the other hand, the
initial rotational energy does constrain the calculated column density
of the rotational levels.  Because the radiative lifetimes of the
higher levels are significantly smaller than that of the lower
rotational levels, a first approximation can be that each level is
populated by its initial population plus the initial population of the
higher level. Therefore, any model yielding an average initial
rotational energy above 0.4 eV would give similar column density
results for $J \leq 7$. Except for \citet{1986MNRAS.223..177D}, most
of the theoretical models agree with a rotationally hot initial
distribution, hence in accordance with our chosen model.

\end{appendix}

\bibliographystyle{apj}
\bibliography{msbib}

\clearpage


\clearpage 

\clearpage   

\clearpage   

\clearpage   

\clearpage   

\clearpage   

\clearpage

\begin{deluxetable}{lcccccc}
\tablecaption{Sightline and Stellar Properties \label{targets}}
\tablewidth{0pt}
\tablehead{
\colhead{Star} & \colhead{l ($^\circ$) }   & \colhead{b ($^\circ$)}   &
\colhead{V (mag)} &
\colhead{E${(B-V)}$ \tablenotemark{a}}  & \colhead{A$_{V}$\tablenotemark{a}} &
\colhead{Sp.T.}
}
\startdata
HD\,73882  & 259.83 & +0.47  & 7.27 &0.72 & 2.28&O9III  \\
HD\,192639  & 74.90 &  +1.48 & 7.11 & 0.66& 1.87&O8e  \\
HD\,206267  & 98.98 & +3.71 & 5.62 & 0.52& 1.37&O6e   \\
HD\,207538  & 102.86 & +6.92 & 7.30 & 0.64 & 1.43&B0V \\
\enddata
\tablenotetext{a}{Extinction parameters from the {\it FUSE} H$_2$ Survey \citep{2002ApJ...577..221R}.}
\end{deluxetable}

\clearpage

\begin{deluxetable}{llcccc}
\tablecaption{Log of {\it FUSE} Observations \label{log}}
\tablewidth{0pt}
\tablehead{
\colhead{Star} & 
\colhead{{\it FUSE} ID\tablenotemark{a}}   & 
\colhead{Observation} &
\colhead{Number of} &
\colhead{Exposure Time} &
\colhead{S/N \tablenotemark{b}} \\
\colhead{} & 
\colhead{} & \colhead{Date}   &
\colhead{Exposures} &
\colhead{(Ks)} &
\colhead{}
}   
\startdata
HD\,73882   & P1161301 & 2000.01.24 & 6 &11.9 & 5.1 \\
..	    & P1161302 & 2000.03.19 & 8 &13.6 & 4.6 \\
HD\,192639  & P1162401 & 2000.06.12 & 2 &4.8 &  8.1 \\
HD\,206267  & P1162701 & 2000.07.21 & 3 &4.9 &  10.2 \\
HD\,207538  & P1162902 & 1999.12.08 & 4 &7.7 &  6.2 \\
... 	    & P1162903 & 2000.07.21 &10 &11.2 & 7.1 \\
 \enddata
\tablenotetext{a}{Archival root name of target for {\it FUSE} PI team observations.}
\tablenotetext{b}{Average per-pixel S/N for a 1 \AA~region of the LIF 1a spectrum near 1070 \AA.}
\end{deluxetable}

\clearpage

\begin{deluxetable}{lcccccc}
\tabletypesize{\scriptsize}
\tablecaption{H$_2$ equivalent width measurements \label{EqW}}
\tablewidth{0pt}
\tablehead{
\colhead{Species} & 
\colhead{$\lambda$ (\AA) } & 
\colhead{log ($f\lambda$)} &
\colhead{HD\,73882  W$_{\lambda}$(m\AA)}  & 
\colhead{HD\,192639  W$_{\lambda}$(m\AA)} & 
\colhead{HD\,206267  W$_{\lambda}$(m\AA)} & 
\colhead{HD\,207538  W$_{\lambda}$(m\AA)} }
\startdata 
H$_{2}$ $J$ = 2.....&       941.606&  0.498&  ...            & 111.4 $\pm$  9.7&  88.3 $\pm$  8.0&  ...            \\
                    &       957.660&  0.661&  ...            & 148.1 $\pm$ 13.5& 115.1 $\pm$  6.1& 147.6 $\pm$ 13.6\\
                    &       975.351&  0.810&  ...            & 151.1 $\pm$ 11.0& 124.9 $\pm$  6.7& 144.7 $\pm$ 10.3\\
                    &       1005.40&  0.998&  ...            & 212.9 $\pm$ 11.5& 162.4 $\pm$  7.7& 208.1 $\pm$ 10.8\\
                    &       1016.47&  1.016& 323.6 $\pm$ 33.3& 245.6 $\pm$ 12.7& 183.2 $\pm$  8.4& 229.1 $\pm$ 10.1\\
                    &       1040.37&  1.030& 359.7 $\pm$ 31.9& 275.7 $\pm$ 13.5& 202.9 $\pm$  8.8& 263.6 $\pm$ 13.5\\
                    &       1053.29&  0.980& 283.0 $\pm$ 23.6& 273.5 $\pm$ 13.1& 193.7 $\pm$  9.3& 270.1 $\pm$ 13.6\\
                    &       1066.91&  0.881& 280.5 $\pm$ 23.9& 257.7 $\pm$ 11.6& 192.9 $\pm$  9.2& 239.7 $\pm$ 10.4\\
                    &       1081.27&  0.709& 256.5 $\pm$ 24.9& 200.1 $\pm$ 11.0& 162.8 $\pm$  6.9& 202.6 $\pm$ 11.4\\
                    &       1096.45&  0.417& 220.1 $\pm$ 27.4& 175.1 $\pm$  8.5&  PF             &  PF             \\
                    &       1112.51& -0.111&  PF             &  PF             &  PF             &  PF             \\
H$_{2}$ $J$ = 3.....&       934.800&  0.820&  ...            & 106.3 $\pm$  9.3&  ...            &  ...            \\
                    &       942.970&  0.729&  ...            & 108.5 $\pm$ 14.2&  85.2 $\pm$  8.5&  ...            \\
                    &       944.337&  0.519&  ...            &  95.2 $\pm$ 13.9&  67.5 $\pm$  9.4&  ...            \\
                    &       958.953&  0.930&  ...            &  95.8 $\pm$  8.6&  97.5 $\pm$  6.5&  90.8 $\pm$  7.7\\
                    &       960.458&  0.674&  ...            &  92.7 $\pm$  8.5&  90.5 $\pm$  6.8&  86.7 $\pm$  5.9\\
                    &       995.974&  1.218&  ...            & 143.9 $\pm$  9.6& 117.6 $\pm$  8.3& 109.2 $\pm$  6.6\\
                    &       997.830&  0.942&  ...            & 123.1 $\pm$ 11.9&  97.5 $\pm$  9.4&  91.4 $\pm$  8.9\\
                    &       1006.42&  1.199&  PF             &  PF             &  PF             &  PF             \\
                    &       1017.43&  1.270&  PF             &  PF             &  PF             &  PF             \\
                    &       1019.51&  1.030& 195.0 $\pm$ 16.4& 116.0 $\pm$  6.9& 113.1 $\pm$  7.2& 101.8 $\pm$  4.5\\
                    &       1028.99&  1.250&  ...            & 144.9 $\pm$ 10.4&  ...            & 128.2 $\pm$  5.3\\
                    &       1041.16&  1.216& 204.0 $\pm$ 12.5& 159.3 $\pm$  9.5& 146.3 $\pm$  7.7& 122.9 $\pm$  6.6\\
                    &       1043.51&  1.052& 182.1 $\pm$ 11.1& 139.5 $\pm$  8.1& 131.1 $\pm$  6.6& 117.6 $\pm$  4.8\\
                    &       1053.98&  1.150& 216.0 $\pm$ 13.4& 159.5 $\pm$ 10.6& 137.6 $\pm$  7.7& 131.7 $\pm$  6.7\\
                    &       1056.48&  1.006& 192.8 $\pm$ 12.0& 125.6 $\pm$  6.7& 122.6 $\pm$  7.5& 121.6 $\pm$  5.5\\
                    &       1067.48&  1.028& 176.0 $\pm$ 15.7& 129.9 $\pm$  7.6& 130.9 $\pm$  7.4& 119.0 $\pm$  5.3\\
                    &       1070.15&  0.909& 196.0 $\pm$ 12.5& 134.6 $\pm$  8.1& 129.0 $\pm$  6.9& 110.0 $\pm$ 10.4\\
                    &       1099.80&  0.448& 148.0 $\pm$ 13.7& 103.1 $\pm$  5.4&  92.1 $\pm$  6.8&  95.2 $\pm$  3.9\\
                    &       1112.59& -0.024&  PF             &  PF             &  PF             &  PF             \\
                    &       1115.91& -0.081& 110.9 $\pm$  9.2&  96.7 $\pm$  6.3&  67.8 $\pm$  3.7&  83.8 $\pm$  4.0\\
H$_{2}$ $J$ = 4.....&       935.969&  1.264&  ...            &  75.2 $\pm$  6.4&  ...            &  ...            \\
                    &       979.808&  1.095&  ...            &  76.7 $\pm$  7.1&  55.2 $\pm$  3.8&  ...            \\
                    &       994.234&  1.134&  ...            &  77.3 $\pm$  8.5&  53.2 $\pm$  6.2&  ...            \\
                    &       999.272&  1.217&  ...            &  88.8 $\pm$  8.1&  54.2 $\pm$  3.3&  59.4 $\pm$  4.9\\
                    &       1017.39&  1.002&  PF             &  PF             &  PF             &  PF             \\
                    &       1032.35&  1.247& 100.7 $\pm$  6.4&  89.0 $\pm$  6.1&  60.9 $\pm$  3.1&  66.2 $\pm$  3.0\\
                    &       1044.55&  1.206&  91.9 $\pm$  4.3&  86.7 $\pm$  4.2&  ...            &  72.9 $\pm$  3.6\\
                    &       1047.56&  1.062&  93.9 $\pm$  6.5&  79.9 $\pm$  4.4&  54.6 $\pm$  2.6&  64.1 $\pm$  3.6\\
                    &       1057.39&  1.135&  97.0 $\pm$  4.5&  72.8 $\pm$  4.2&  59.9 $\pm$  4.1&  65.8 $\pm$  3.2\\
                    &       1060.59&  1.019&  78.3 $\pm$  5.3&  81.2 $\pm$  3.9&  57.1 $\pm$  2.8&  63.0 $\pm$  3.4\\
                    &       1074.32&  0.923&  91.1 $\pm$  5.6&  71.2 $\pm$  3.7&  51.5 $\pm$  3.3&  57.7 $\pm$  5.2\\
                    &       1085.15&  0.817&  ...            &  70.5 $\pm$  8.8&  45.3 $\pm$  4.0&  ...            \\
                    &       1088.80&  0.752&  78.8 $\pm$  5.4&  71.6 $\pm$  4.0&  46.7 $\pm$  3.1&  58.4 $\pm$  2.8\\
                    &       1100.17&  0.498&  62.5 $\pm$  4.2&  73.2 $\pm$  5.5&  43.6 $\pm$  3.7&  ...            \\
                    &       1104.09&  0.461&  ...            &  64.5 $\pm$  3.8&  ...            &  ...            \\
                    &       1116.03& -0.060&  34.1 $\pm$  4.5&  44.7 $\pm$  4.8&  17.6 $\pm$  2.9&  28.3 $\pm$  3.4\\
                    &       1120.26& -0.069&  30.5 $\pm$  3.0&  45.4 $\pm$  5.9&  20.2 $\pm$  2.5&  33.3 $\pm$  3.1\\
H$_{2}$ $J$ = 5.....&       942.691&  0.765&  ...            &  37.5 $\pm$  8.7&  ...            &  ...            \\
                    &       974.889&  1.138&  ...            &  68.4 $\pm$  6.3&  38.8 $\pm$  3.8&  ...            \\
                    &       996.129&  1.102&  ...            &  PF             &  45.4 $\pm$  6.3&  40.2 $\pm$  8.0\\
                    &       997.644&  1.110&  ...            &  62.3 $\pm$  6.4&  44.5 $\pm$  4.8&  33.3 $\pm$ 10.1\\
                    &       1006.34&  0.940&  PF             &  PF             &  PF             &  PF             \\
                    &       1017.01&  1.060&  88.3 $\pm$  9.0&  67.6 $\pm$  4.5&  PF             &  PF             \\
                    &       1017.84&  1.384& 117.6 $\pm$  7.5&  82.0 $\pm$  3.8&  51.4 $\pm$  3.1&  56.4 $\pm$  5.0\\
                    &       1040.06&  1.074&  91.2 $\pm$  5.4&  64.9 $\pm$  3.4&  37.1 $\pm$  2.7&  33.7 $\pm$  3.6\\
                    &       1052.50&  1.068&  80.8 $\pm$  6.2&  64.8 $\pm$  3.2&  38.4 $\pm$  3.4&  30.7 $\pm$  3.0\\
                    &       1061.70&  1.126&  87.8 $\pm$ 12.0&  76.8 $\pm$  3.4&  39.1 $\pm$  2.5&  45.4 $\pm$  4.3\\
                    &       1065.60&  1.026&  93.5 $\pm$  9.2&  67.9 $\pm$  3.6&  33.0 $\pm$  3.2&  27.8 $\pm$  3.2\\
                    &       1075.25&  0.999&  94.9 $\pm$  6.9&  70.6 $\pm$  2.9&  36.6 $\pm$  2.8&  31.6 $\pm$  3.5\\
                    &       1089.52&  0.796&  68.9 $\pm$  4.3&  58.1 $\pm$  4.6&  31.2 $\pm$  3.4&  ...            \\
                    &       1104.55&  0.471&  ...            &  30.9 $\pm$  3.0&  ...            &  ...            \\
                    &       1109.32&  0.467&  47.3 $\pm$  4.1&  31.4 $\pm$  3.1&  16.6 $\pm$  2.0&  12.4 $\pm$  2.2\\
                    &       1120.41& -0.095&  19.3 $\pm$  3.3&  14.7 $\pm$  2.7&   5.0 $\pm$  0.9&  PF             \\
H$_{2}$ $J$ = 6.....&       959.163&  1.426&  ...            &  16.4 $\pm$  4.2&  ...            &  ...            \\
                    &       977.733&  1.538&  ...            &  20.1 $\pm$  4.2&  ...            &  ...            \\
                    &       998.340&  1.564&  ...            &  22.6 $\pm$  4.2&  ...            &  ...            \\
                    &       1019.02&  1.318&  31.4 $\pm$  4.9&  14.0 $\pm$  2.4&  ...            &  ...            \\
                    &       1021.22&  1.388&  22.7 $\pm$  6.9&  20.2 $\pm$  3.6&  ...            &  ...            \\
                    &       1041.74&  1.243&  26.1 $\pm$  3.9&  11.8 $\pm$  2.9&  ...            &  ...            \\
                    &       1045.81&  1.076&  16.1 $\pm$  5.1&  15.4 $\pm$  3.5&  ...            &  ...            \\
                    &       1058.32&  1.074&  18.0 $\pm$  3.8&  ...            &  ...            &  ...            \\
H$_{2}$ $J$ = 7.....&       1060.04&  1.193&  20.0 $\pm$  3.9&  ...            &  ...            &  ...            \\
                    &       1073.00&  1.110&  14.2 $\pm$  3.2&  ...            &  ...            &  ...            \\
\enddata
\tablecomments{Lines blended but used for profile fitting are noted ``PF''. Errors are 1 $\sigma$.}
\end{deluxetable}

\clearpage

\begin{deluxetable}{lcccc}
\tablecaption{COG determined $b$ values for different combinations
of EqW towards HD\,192639\label{COGtest}}
\tabletypesize{\scriptsize}
\tablewidth{0pt}
\tablehead{
\colhead{Level} & \colhead{All the EqW} &
\colhead{Half of the EqW}   & \colhead{Second half of the EqW} 
}   
\startdata
H$_{2}$ $J$ = 2 &  $  3.8 \pm~^{1.3}_{2.3} $ & $ 4.3 \pm~^{2.1}_{2.6} $ & $ 3.3 \pm~^{2.1}_{2.6} $  \\ 
H$_{2}$ $J$ = 3 &  $  4.8 \pm~^{0.8}_{0.7} $ & $ 5.0 \pm~^{1.2}_{0.8} $ & $ 4.5 \pm~^{1.6}_{1.6} $  \\ 
H$_{2}$ $J$ = 4 &  $  5.6 \pm~^{0.6}_{0.5} $ & $ 5.4 \pm~^{0.6}_{1.1} $ & $ 6.0 \pm~^{0.7}_{0.8} $  \\ 
H$_{2}$ $J$ = 5 &  $  8.0 \pm~^{1.0}_{0.9} $ & $ 7.8 \pm~^{2.2}_{1.4} $ & $ 7.9 \pm~^{1.4}_{1.0} $  \\ 
\enddata
\tablecomments{$b$ values are in km\,s$^{-1}$. Errors are 2 $\sigma$. }
\end{deluxetable}

\clearpage

\begin{deluxetable}{lcccccccc}
\tablecaption{Log column densities and $b$ values obtained with PF and COG for the four
lines of sight \label{tb:res}}
\tabletypesize{\scriptsize}
\tablewidth{0pt}
\tablehead{
\colhead{ {\rm $J$ level} } &
\multicolumn{2}{c}{HD\,73882} &
\multicolumn{2}{c}{HD\,192639} &
\multicolumn{2}{c}{HD\,206267} &
\multicolumn{2}{c}{HD\,207538} \\
\colhead{cm$^{-2}$ / km\,s$^{-1}$} & 
\colhead{ COG } & \colhead{ PF } &
\colhead{ COG } & \colhead{ PF } &
\colhead{ COG } & \colhead{ PF } &
\colhead{ COG } & \colhead{ PF } 
}
\startdata
$N({\rm H}_2)$ --- $J = 0$ &\multicolumn{2}{c}{$20.99 \pm~^{0.08}_{0.08}$}&\multicolumn{2}{c}{$20.28 \pm~^{0.05}_{0.05}$}&\multicolumn{2}{c}{$20.64 \pm~^{0.03}_{0.03}$}&\multicolumn{2}{c}{$20.64 \pm~^{0.07}_{0.07}$} \\
$N({\rm H}_2)$ --- $J = 1$ &\multicolumn{2}{c}{$20.50 \pm~^{0.07}_{0.07}$}&\multicolumn{2}{c}{$20.48 \pm~^{0.05}_{0.05}$}&\multicolumn{2}{c}{$20.45 \pm~^{0.05}_{0.05}$}&\multicolumn{2}{c}{$20.58 \pm~^{0.05}_{0.05}$} \\
$N({\rm H}_2)$ --- $J = 2$ &$18.76 \pm~^{0.14}_{0.54}$&$19.04 \pm~^{0.10}_{0.09}$&$18.66 \pm~^{0.05}_{0.05}$&$18.75 \pm~^{0.10}_{0.05}$&$18.39 \pm~^{0.06}_{0.06}$&$18.49 \pm~^{0.05}_{0.04}$&$18.63 \pm~^{0.05}_{0.07}$&$18.75 \pm~^{0.05}_{0.05}$ \\
$N({\rm H}_2)$ --- $J = 3$ &$18.22 \pm~^{0.10}_{0.15}$&$18.49 \pm~^{0.15}_{0.04}$&$17.84 \pm~^{0.11}_{0.16}$&$18.00 \pm~^{0.15}_{0.10}$&$17.91 \pm~^{0.06}_{0.08}$&$17.95 \pm~^{1.05}_{0.05}$&$17.65 \pm~^{0.11}_{0.16}$&$17.90 \pm~^{0.10}_{0.15}$ \\
$N({\rm H}_2)$ --- $J = 4$ &$15.66 \pm~^{0.10}_{0.09}$&$15.80 \pm~^{0.10}_{0.16}$&$16.01 \pm~^{0.20}_{0.17}$&$15.90 \pm~^{0.10}_{0.10}$&$15.46 \pm~^{0.11}_{0.12}$&$15.28 \pm~^{0.17}_{0.10}$&$15.77 \pm~^{0.12}_{0.14}$&$15.70 \pm~^{0.10}_{0.06}$ \\
$N({\rm H}_2)$ --- $J = 5$ &$15.34 \pm~^{0.11}_{0.10}$&$15.34 \pm~^{0.10}_{0.11}$&$15.18 \pm~^{0.08}_{0.07}$&$15.15 \pm~^{0.08}_{0.07}$&$14.83 \pm~^{0.10}_{0.10}$&$14.75 \pm~^{0.15}_{0.10}$&$14.59 \pm~^{0.13}_{0.06}$&$14.64 \pm~^{0.16}_{0.15}$ \\
$N({\rm H}_2)$ --- $J = 6$ &$14.28 \pm~^{0.11}_{0.14}$&$14.08 \pm~^{0.26}_{0.08}$&$13.98 \pm~^{0.11}_{0.13}$&$14.00 \pm~^{0.15}_{0.10}$& \nodata &$13.84 \pm~^{0.15}_{?}$& \nodata &$13.04 \pm~^{0.60}_{?}$\\
$N({\rm H}_2)$ --- $J = 7$ &$14.16 \pm~^{0.17}_{0.22}$&$14.08 \pm~^{0.37}_{0.23}$& \nodata &$13.49 \pm~^{0.35}_{?}$& \nodata
&$13.59 \pm~^{0.31}_{?}$& \nodata &$13.28 \pm~^{0.56}_{?}$ \\
\\
\tableline
\\
$b ({\rm H}_2)$ --- $J = 2$ & \nodata &$2.47 \pm~^{3.0}_{1.7}$ &$  3.8 \pm~^{1.3}_{2.3} $ & $ 4.59 \pm~^{0.40}_{0.70}$  &$  3.8 \pm~^{0.9}_{1.3} $ &$ 2.49 \pm~^{0.20}_{0.40}$  & $  3.3 \pm~^{2.3}_{3.3} $ & $ 3.97 \pm~^{0.20}_{0.20}$  \\
$b ({\rm H}_2)$ --- $J = 3$ &   $  5.7 \pm~^{1.3}_{1.3} $ & $4.39 \pm~^{0.4}_{0.2}$&  $  4.8 \pm~^{0.8}_{0.7} $ & $ 5.14 \pm~^{0.20}_{0.20}$  &    $  3.5 \pm~^{0.5}_{0.5} $ &  $ 3.10 \pm~^{0.30}_{0.10}$  &  $  4.5 \pm~^{0.6}_{0.5} $ & $ 4.41 \pm~^{0.20}_{0.20}$ \\
$b ({\rm H}_2)$ --- $J = 4$ &  $  7.7 \pm~^{0.7}_{0.7} $ &$7.19 \pm~^{0.4}_{0.3}$  &  $  5.6 \pm~^{0.6}_{0.5} $ &$ 5.89 \pm~^{0.20}_{0.30}$  & $  4.7 \pm~^{0.5}_{0.4} $ &$ 5.02 \pm~^{0.15}_{0.45}$  & $  4.8 \pm~^{0.5}_{0.4} $ &$ 4.78 \pm~^{0.20}_{0.30}$  \\
$b ({\rm H}_2)$ --- $J = 5$ &  $ 10.1 \pm~^{1.9}_{1.5} $  &  $11.09 \pm~^{1.2}_{1.0}$  &  $  8.0 \pm~^{1.0}_{0.9} $ &$ 8.34 \pm~^{0.80}_{0.40}$ & $  5.3 \pm~^{1.5}_{0.9} $ & $ 6.50 \pm~^{0.60}_{1.40}$  &   $ 10.8 \pm~^{1.2}_{5.1} $ & $ 7.22 \pm~^{2.40}_{1.60}$ \\
\enddata
\tablecomments{$J =$ 0 \& 1 column densities from \cite{2002ApJ...577..221R}. Errors are 2 $\sigma$.} 
\end{deluxetable}
\clearpage

\begin{deluxetable}{lccc}
\tablecaption{Gas Densities and H$_2$ Formation rates --- $b$ due to H$_2$-formation pumping\label{tb:resmodel}}
\tablewidth{0pt}
\tablehead{
\colhead{} &
\colhead{$n$}& 
\colhead{$nR$}&
\colhead{$R$}\\
\colhead{Lines of Sight} &
\colhead{[cm$^{-3}$]}& 
\colhead{[s$^{-1}$]}&
\colhead{[cm$^3$\,s$^{-1}$]}
}
\startdata
{HD\,73882  }   & $5 \pm~^{3}_{?}$   & $9 \pm~^{8}_{4}\times 10^{-15}$ & $1.8 \pm~^{?}_{1.2}\times 10^{-15}$\\
{HD\,192639  }  & $13 \pm~^{5}_{6}$  & $4 \pm~^{5}_{3}\times 10^{-15}$ & $3.1 \pm~^{9.8}_{2.5}\times 10^{-16}$\\
{HD\,206267  }  & $25 \pm~^{11}_{6}$ & $6 \pm~^{4}_{3}\times 10^{-15}$ & $2.4 \pm~^{2.9}_{2.1}\times 10^{-16}$\\
{HD\,207538  }  & $13 \pm~^{20}_{?}$ & $3 \pm~^{2}_{2}\times 10^{-15}$ & $2.3 \pm~^{?}_{1.9}\times 10^{-16}$\\
\enddata
\tablecomments{Densities and formation rates come from our
  measurements reported in Table~\ref{tb:res} and model results
  plotted in Figure~\ref{modeleps}. }
\end{deluxetable}

\clearpage

\begin{deluxetable}{lcccc}
\tablecaption{Temperatures and Densities --- $b$ due to thermal scattering\label{tb:warm}}
\tablewidth{0pt}
\tablehead{
\colhead{} &
\colhead{$T^{k}$}& 
\colhead{$T^{ex}_{J=5-7}$}& 
\colhead{$n$}&
\colhead{$log_{10}(P/k)$}\\
\colhead{ Lines of Sight} &
\colhead{(K)}& 
\colhead{(K)}& 
\colhead{(cm$^{-3}$)}&
\colhead{(log(K cm$^{-3}$))}
}
\startdata
{HD\,73882  }   & $9500 \pm~^{600}_{600} $&$ 670 \pm~^{180}_{100} $&$31 \pm~^{30}_{14}$&$4.3 \pm~^{0.4}_{0.3}$\\
{HD\,192639  }  & $8200 \pm~^{400}_{300} $&$ 500 \pm~^{120}_{?} $&$13 \pm~^{17}_{?}$&$3.8 \pm~^{0.5}_{?}$\\
{HD\,206267  }  & $7000 \pm~^{600}_{700} $&$ 680 \pm~^{220}_{?} $&$49 \pm~^{54}_{?}$&$4.5 \pm~^{0.4}_{?}$\\
{HD\,207538  }  & $8500 \pm~^{800}_{1700}$&$ 610 \pm~^{400}_{?} $&$27 \pm~^{75}_{?}$&$4.2 \pm~^{0.8}_{?}$\\
\enddata
\tablecomments{$T^{k}$ is the temperature obtained from the formula $b=\sqrt{2kT/m}$.
  $T^{ex}_{J=5-7}$ is the excitation temperature from the
  rotational levels 5 and 7.}
\end{deluxetable}




\end{document}